\documentclass[prb,twocolumn,showkeys,showpacs,preprintnumbers,amsmath,amssymb,floatfix]{revtex4}

\usepackage{graphicx}
\usepackage{dcolumn}
\usepackage{bm}
\usepackage{ulem}

\begin{document}

\title{Magnetic anisotropy in (Ga,Mn)As: Influence of epitaxial strain and hole concentration}

\author{M. Glunk}
\author{J. Daeubler}
\author{L. Dreher}
\author{S. Schwaiger}
\author{W. Schoch}
\author{R. Sauer}
\author{W. Limmer}
\email{wolfgang.limmer@uni-ulm.de}
\affiliation{Institut f\"ur Halbleiterphysik, Universit\"at Ulm,
89069 Ulm, Germany}
\author{A. Brandlmaier}
\author{S. T. B. Goennenwein}
\affiliation{Walther-Meissner-Institut, Bayerische Akademie der
Wissenschaften, Walther-Meissner-Strasse 8, 85748 Garching, Germany}
\author{C. Bihler}
\author{M. S. Brandt}
\affiliation{Walter Schottky Institut, Technische Universit\"at
M\"unchen, Am Coulombwall 3, 85748 Garching, Germany}

% \date{\today}

\begin{abstract}
We present a systematic study on the influence of epitaxial strain and
hole concentration on the magnetic anisotropy in (Ga,Mn)As at 4.2\;K.
The strain was gradually varied over a wide range from tensile to
compressive by growing a series of (Ga,Mn)As layers with 5\% Mn on
relaxed graded (In,Ga)As/GaAs templates with different In concentration.
The hole density, the Curie temperature, and the relaxed lattice constant
of the as-grown and annealed (Ga,Mn)As layers turned out to be essentially
unaffected by the strain. Angle-dependent magnetotransport measurements
performed at different magnetic field strengths were used to probe the
magnetic anisotropy. The measurements reveal a pronounced linear dependence
of the uniaxial out-of-plane anisotropy on both strain and hole density.
Whereas the uniaxial and cubic in-plane anisotropies are nearly constant,
the cubic out-of-plane anisotropy changes sign when the magnetic easy
axis flips from in-plane to out-of-plane. The experimental results for
the magnetic anisotropy are quantitatively compared with calculations
of the free energy based on a mean-field Zener model. An almost perfect
agreement between experiment and theory is found for the uniaxial
out-of-plane and cubic in-plane anisotropy parameters of the as-grown
samples. In addition, magnetostriction constants are derived from the
anisotropy data.
\end{abstract}

\pacs{75.50.Pp, 75.30.Gw, 75.47.--m, 61.05.--a}

\keywords{(Ga,Mn)As; (In,Ga)As; Strain; Magnetic anisotropy;
Magnetotransport}

\maketitle

\section{\label{introduction}Introduction}
Spin-related phenomena in semiconductors, such as spin polarization,
magnetic anisotropy (MA), and anisotropic magnetoresistance (AMR),
open up new concepts for information processing and storage beyond
conventional electronics.\cite{Zut04,Pea05} Being compatible with
the standard semiconductor GaAs, the dilute magnetic semiconductor
(Ga,Mn)As has proven to be an ideal playground for studying future
spintronic applications.\cite{Mac05,Jun06} In particular, the
pronounced MA and AMR, largely arising from the spin-orbit coupling
in the valence band,\cite{Die01,Abo01} potentially apply in novel
non-volatile memories and magnetic-field-sensitive devices.
Ferromagnetism is implemented in (Ga,Mn)As by incorporating high
concentrations ($\gtrsim$1\%) of magnetic Mn$^{2+}$ ions into the Ga
sublattice. The ferromagnetic coupling between the S=5/2 Mn spins is
mediated by itinerant holes provided by the Mn acceptor itself. Curie
temperatures $T_{\text{C}}$ up to 185\;K, i.e. well above the
liquid-N$_2$ temperature, have been reported\cite{Wan08,Nov08} and
there is no evidence for a fundamental limit to higher
values.\cite{Jun05}

The magnetic properties of (Ga,Mn)As are strongly temperature
dependent and can be manipulated to a great extent by doping,
material composition, and strain. In (Ga,Mn)As grown on GaAs
substrates, however, hole density $p$, Mn concentration $x$, and
strain $\varepsilon$ are intimately linked to each other and cannot
be tuned independently by simply varying the growth parameters.
$p$ and $\varepsilon$ sensitively depend on the concentration and
distribution of the Mn atoms which are incorporated both on Ga
lattice sites (Mn$_{\text{Ga}}$) and, to a lower extent, on
interstitial sites (Mn$_{\text{I}}$), where they act as compensating
double donors. Post-growth treatment techniques such as annealing
or hydrogenation are frequently used to increase or decrease the
hole concentration due to outdiffusion and/or rearrangement of
Mn$_{\text{I}}$\cite{Pot01,Yu02,Edm04,Zha05,Lim05} or due to the
formation of electrically inactive (Mn,H)
complexes,\cite{Goe04,The07,Bih08c} respectively. In both cases,
however, the treatment concurrently leads to a decrease or
increase of the lattice parameter, respectively, and thus to a
change of the strain.

The epitaxial strain in the (Ga,Mn)As layers, arising from
the lattice mismatch between layer and substrate, can be
adjusted by tailoring the lattice parameter of the substrate.
While (Ga,Mn)As grown on GaAs is under compressive strain,
tensily strained (Ga,Mn)As can be obtained by using appropriate
(In,Ga)As/GaAs templates.\cite{Jun06,She97,Liu03,The06}
Experimental studies addressing this issue, however, have so far
been restricted to merely a limited number of representative
samples.

In this work, the influence of epitaxial strain and hole
concentration on the MA at 4.2\;K is analyzed in a systematic way
by investigating a set of (Ga,Mn)As layers grown on relaxed
(In,Ga)As/GaAs templates with different In concentration. Keeping
the Mn content at $\sim$5\% and changing the maximum In content in
the (In,Ga)As buffer layers from 0\% to 12\%, the vertical strain
$\varepsilon_{zz}$ in the as-grown (Ga,Mn)As layers could be
gradually varied over a wide range from $\varepsilon_{zz}$=0.22\%
in the most compressively strained sample to
$\varepsilon_{zz}$=$-0.38$\% in the most tensily strained sample
without substantially changing $p$. Post-growth annealing leads to
an increase in $p$, yielding a second series of samples with nearly
the same range of $\varepsilon_{zz}$ but higher hole concentrations.
The strain dependence of the anisotropy parameters for the as-grown
and the annealed samples was determined by means of angle-dependent
magnetotransport measurements.\cite{Lim06a, Lim08}
Part of the experimental data has already been published in
conference proceedings.\cite{Dae08} Here, we combine the earlier
with the present extensive experimental findings advancing a
quantitative comparison of the intrinsic anisotropy parameters
with model calculations for the MA, performed within the mean-field
Zener model introduced by Dietl et al.\cite{Die01}
Note that several samples analyzed in Ref.~\onlinecite{Dae08} have
been substituted by new samples grown under optimized
conditions and that the sample series has been expanded
by one specimen with $\varepsilon_{zz}$=-0.38\%.

\section{\label{experimental}Experimental details}

A set of differently strained (Ga,Mn)As layers with constant Mn
concentration of $\sim$5\% and thickness of $\sim$180\;nm was grown
by low-temperature molecular-beam epitaxy (LT-MBE) on (In,Ga)As/GaAs
templates with different In content in a RIBER 32 MBE machine.
Indium-mounted semi-insulating VGF GaAs(001) wafers were used as
substrates. After thermal deoxidation, a 30-nm-thick GaAs buffer
layer was deposited at a substrate temperature of
$T_{\text{s}}$$\approx$580\;$^{\circ}$C. Then the growth was
interrupted, $T_{\text{s}}$ was lowered to $\sim$430\;$^{\circ}$C,
and a graded (In,Ga)As buffer with a total thickness of up to
$\sim$5\;$\mu$m was grown. Starting with In$_{0.02}$Ga$_{0.98}$As,
the temperature of the In cell was first continuously raised to
increase the In content up to a value of $\leq$12\% and was then
kept constant until the required thickness of the buffer layer was
reached. The growth was again interrupted, $T_{\text{s}}$ was
lowered to $\sim$250\;$^{\circ}$C, and the (Ga,Mn)As layer was
grown in As$_4$ mode at a growth rate of $\sim$200\;nm/h.
The growth was monitored by reflection high-energy electron
diffraction showing no indication of a second-phase formation.
The use of a graded (In,Ga)As buffer\cite{Har89} minimizes the
deterioration of the (Ga,Mn)As layer caused by threading
dislocations in the relaxed (In,Ga)As/GaAs template. The resulting
(Ga,Mn)As layers exhibit nearly the same quality as conventional
samples directly grown on GaAs.\cite{The06} After the growth, the
samples were cleaved into several pieces and some of the pieces
were annealed in air for 1\;h at 250\;$^{\circ}$C. The structural
properties of the (Ga,Mn)As layers were analyzed by means of
high-resolution x-ray diffraction (HRXRD) measurements performed
with a Siemens D5000HR x-ray diffractometer using the
Cu-K$_{\alpha_1}$ radiation at 0.154\;nm. Hall bars with current
directions along the $[100]$ and $[110]$ crystallographic axes
were prepared from the samples by standard photolithography and
wet chemical etching. The width of the Hall bars is 0.3\;mm and
the longitudinal voltage probes are separated by 1\;mm. The hole
densities were determined by high-field magnetotransport
measurements (up to 14.5\;T) at 4.2\;K using an Oxford SMD 10/15/9
VS liquid-helium cryostat with superconducting coils. The Curie
temperatures were estimated from the peak positions of the
temperature-dependent sheet resistivities at
10\;mT.\cite{Nov08,Esc97,Mat98}
The MA of the samples was probed by means of
angle-dependent magnetotransport measurements at 4.2\;K using a
liquid-He bath cryostat equipped with a rotatable sample holder and
a standard LakeShore electromagnet system with a maximum field
strength of 0.68\;T. To determine the saturation magnetization, we
employed a Quantum Design MPMS-XL-7 superconducting quantum
interference device (SQUID) magnetometer using the Reciprocating
Sample Option (RSO). The measured SQUID curves were corrected for
the diamagnetic contribution of the substrate.

As discussed in detail in Ref.~\onlinecite{Bih08b}, our samples
exhibit spin wave resonances which are most pronounced for the
external magnetic field oriented perpendicular to the sample
plane. These spin wave excitations have been traced back to an
inhomogeneous free-energy density profile, or more precisely to
a linear variation of the MA parameters along the growth direction,
presumably arising from a vertical gradient in the hole
density.\cite{Koe03,Lim05} Therefore, all physical parameters
derived via magnetotransport in this study have to be considered as
\textit{effective} parameters representing the averaged electronic
and magnetic properties of the layers.

\section{Theoretical considerations \label{theory}}

In the present context, MA represents the dependence of the
free-energy density $F$ on the orientation $\bm{m}$ of the
magnetization $\bm{M}=M \bm{m}$.\cite{rem} In the absence of an
external magnetic field, $\bm{m}$ is determined by the minimum of
the free energy. Since for reasons of crystal symmetry $F$ usually
exhibits several equivalent minima, more than one stable orientation
of $\bm{M}$ exists. This symmetry-induced degeneracy of $F$ can be
lifted by the application of an external magnetic field $\bm{H}$.

The theoretical considerations on the AMR and the MA in this paper
are based on a single-domain model with a uniform magnetization
$\bm{M}$. While the direction of $\bm{M}$ is controlled by the
interplay of $F$ and $\bm{H}$, its magnitude $M$ is assumend
to be constant under the given experimental conditions. For
sufficiently high field strengths $H$, this assumption can be
considered as a good approximation. Hence, the normalized quantity
$F_M$=$F/M$ is considered instead of $F$, allowing for a concise
description of the MA.

\subsection{Phenomenological description of the MA \label{macroscopic_model}}

There are several contributions to $F_M$ which we refer to as intrinsic
(magnetocrystalline) or extrinsic:
\begin{equation}
F_M = F_{M,\text{int}}+F_{M,\text{ext}}\;.
\label{F_M}
\end{equation}
The intrinsic part $F_{M,\text{int}}$=$F_{M,\text{c}}+F_{M,\text{S}}$
originates from the holes in the valence band ($F_{M,\text{c}}$)
(see Sec.~\ref{comparison}) and from the localized Mn spins
($F_{M,\text{S}}$).\cite{Die01} Whereas $F_{M,\text{c}}$ is strongly
anisotropic with respect to the magnetization orientation, reflecting
the anisotropy of the valence band, the localized-spin contribution
$F_{M,\text{S}}$=$\int_0^MdM'\mu_0H(M')/M$ is isotropic and therefore
irrelevant for the following considerations. In a
phenomenological description, $F_{M,\text{int}}$ can be expressed in
terms of a series expansion in ascending powers of the direction
cosines $m_x$, $m_y$, and $m_z$ of the magnetization with respect to
the cubic axes [100], [010], and [001], respectively. Considering
terms up to the fourth order in $\bm{m}$, $F_{M,\text{int}}$ for
cubic systems with tetragonal distortion along the [001] growth
direction is given by\cite{Lim08}
\begin{equation}
F_{M,\text{int}}(\bm{m}) = B_0 + B_{2\perp}m_z^2 +
B_{4\parallel}(m_x^4+m_y^4) + B_{4\perp}m_z^4\;.
\label{F_MA_int}
\end{equation}
In the case of a perfect cubic crystal, symmetry requires
$B_{2\perp}$=0 and $B_{4\parallel}$=$B_{4\perp}$.

The extrinsic part $F_{M,\text{ext}}$ comprises the demagnetization
energy due to shape anisotropy and a
uniaxial in-plane anisotropy along [$\bar{1}10$]. The origin of
the latter anisotropy is controversially discussed. It is traced back
either to highly hole-concentrated (Ga,Mn)As clusters formed during
the growth,\cite{Ham06} to the anisotropy of the reconstructed
initial GaAs (001) substrate surface,\cite{Wel04} or to a trigonal-like
distortion which may result from a nonisotropic Mn distribution,
caused, for instance, by the presence of surface dimers oriented along
[$\bar{1}10$] during the epitaxy.\cite{Saw05}
Approximating the (Ga,Mn)As layer by an infinite plane, we write
the total extrinsic contribution as
\begin{equation}
F_{M,\text{ext}}(\bm{m}) = B_{\text{d}} m_z^2 +
B_{\bar110}\frac{1}{2}(m_x-m_y)^2,
\label{F_MA_ext}
\end{equation}
where $B_{\text{d}}$=$\mu_0M/2$.

In the presence of an external magnetic field $\bm{H}$=$H\bm{h}$, the
normalized Zeeman energy $-\mu_0\bm{H}\bm{m}$ has to be added to the
total free-energy density. This corresponds to a transition from $F_M$
to the normalized free-enthalpy density
\begin{eqnarray}
G_M(\bm{m}) &=& B_0 + (\overbrace{B_{2\perp}+B_{\text{d}}}^{B_{001}})m_z^2
\nonumber \\
&& + B_{4\parallel}(m_x^4+m_y^4)+ B_{4\perp}m_z^4 \nonumber \\
&& + B_{\bar110}\frac{1}{2}(m_x-m_y)^2 - \mu_0H\bm{h}\bm{m}. \label{G_MA}
\end{eqnarray}
The anisotropy parameters $B_{2\perp}$ and $B_{\text{d}}$ are both
related to $m_z^2$ and are therefore combined into a single parameter
$B_{001}$. Given an arbitrary magnitude and orientation of $\bm{H}$,
the direction of $\bm{m}$ is determined by the minimum of $G_M$.

All anisotropy parameters introduced above are in SI units.
Expressed by the anisotropy fields in cgs units as defined, e.g., in
Ref.~\onlinecite{Liu06}, they read as
$B_{\bar{1}10}$=$-\mu_0H_{2\parallel}/2$, $B_{2\perp}$=$-\mu_0H_{2\perp}/2$,
$B_{4\parallel}$=$-\mu_0H_{4\parallel}/4$, and $B_{4\perp}$=$-\mu_0H_{4\perp}/4$.
Note also that the magnetic anisotropy field used in
Ref.~\onlinecite{Bih08b} and the anisotropy parameters used here are related via
$\mu_0H^{001}_{\rm aniso}=2(K^{001}_{\rm eff}+ K^{\perp}_{\rm c1})/M_{\rm sat}=-2B_{001}-4B_{4\perp}$.

\subsection{Microscopic theory \label{comparison}}

For a microscopic description of the intrinsic part $F_{M,\text{int}}$,
we adopt the mean-field Zener model of Dietl et al. introduced in
Ref.~\onlinecite{Die01}. The objective of the microscopic calculations
discussed below is first, to justify the approximation in
Eq.~(\ref{F_MA_int}), made by considering only terms up to the fourth
order, and second, to compare the experimentally found dependence of the
intrinsic anisotropy parameters $B_{2\perp}$, $B_{4\parallel}$, and
$B_{4\perp}$ on $\varepsilon_{zz}$ and $p$ (see
Sec.~\ref{anisotropy_parameters}) with that predicted by the mean-field
Zener model.

According to the $\bm{k}\cdot\bm{p}$ effective Hamiltonian theory
presented in Ref.~\onlinecite{Die01}, the Hamiltonian of the
system is given by
\begin{equation}
\mathcal{H}=\mathcal{H}_{\text{KL}}+\mathcal{H}_{\varepsilon}+
\mathcal{H}_{\text{pd}}.
\label{hamiltonian}
\end{equation}
Here, $\mathcal{H}_{\text{KL}}$ represents the 6$\times$6
Kohn-Luttinger \mbox{$\bm{k}\cdot\bm{p}$} Hamiltonian for the
valence band and
$\mathcal{H}_{\varepsilon}$=$\sum_{i,j}D^{(ij)}\varepsilon_{ij}$
accounts for the strain $\varepsilon_{ij}$ in the (Ga,Mn)As layer
via the deformation potential operator $D^{(ij)}$.
$\mathcal{H}_{\text{pd}}$=$-N_0\beta\bm{S}\bm{s}$ describes the
p-d hybridization of the p-like holes and the localized Mn d-shell
electrons, which results in an interaction between the hole spin
$\bm{s}$ and the Mn spin $\bm{S}$ carrying a magnetic moment
$Sg\mu_{\text{B}}$. Here, $g$=2 is the Land\'{e} factor,
$\mu_{\text{B}}$ the Bohr magneton, and $\beta$ and $N_0$ denote
the p-d exchange integral and the concentration of cation sites,
respectively. In terms of the virtual crystal and mean-field
approximation, the exchange interaction can be written as
$\mathcal{H}_{\text{pd}}$=$\bm{M}\bm{s}\beta/g\mu_{\text{B}}$.
Explicit expressions for the individual contributions in
Eq.~(\ref{hamiltonian}) can be found in Ref.~\onlinecite{Die01}.
As an approximation, the values of the Luttinger parameters
$\gamma_i$ ($i$=1,2,3), the spin-orbit splitting $\Delta_0$,
and the valence band shear deformation potential $b$ are chosen
as those of GaAs. Explicit values are $\gamma_1$=6.85, $\gamma_2$=2.1,
$\gamma_3$=2.9, $\Delta_0$=0.34\;eV and $b$=$-1.7$\;eV,
respectively.\cite{Die01} The quantity parameterizing the
exchange splitting of the valence subbands is given by
\begin{equation}
B_G=\frac{A_\text{F}\beta M}{6g\mu_{\rm B}}\,, \label{B_G}
\end{equation}
with the Fermi liquid parameter $A_{\text{F}}$. In contrast
to Ref.~\onlinecite{Die01}, we restrict our calculations
to zero temperature ($T$=0) and zero magnetic field ($H$=0).
In this approximation, the Fermi distribution is represented by a
step function and the Zeeman as well as the Landau splitting can be
neglected. These simplifications are justified, as our measurements
were carried out at $T$=4.2\;K and $\mu_0H<0.7$\;T, where the
Zeeman and Landau splittings of the valence band are expected to be
much smaller than the splitting caused by the p-d exchange coupling.
Diagonalization of the Hamilton matrix $\mathcal{H}$ yields the
sixfold spin-split valence band structure in the vicinity of the
$\Gamma$ point, depending on the magnetization orientation
$\bm{m}$ and the strain $\varepsilon_{ij}$ in the (Ga,Mn)As layer.

The $\bm{m}$- and $\varepsilon_{ij}$-dependent normalized free-energy
density of the carrier system $F_{M,\text{c}}(\varepsilon_{ij},\bm{m})$
is obtained by first summing over all energy eigenvalues within the
four spin-split heavy-hole and light-hole Fermi surfaces and then
dividing the resulting energy density by $M$. The two split-off
valence bands do not contribute to $F_{M,\text{c}}$ because they
lie energetically below the Fermi energy for common carrier
concentrations. For biaxially strained (Ga,Mn)As layers grown
pseudomorphically on (001)-oriented substrates, the tetragonal
distortion of the crystal lattice along [001] can be fully described
by the $\varepsilon_{zz}$ component of the strain tensor using
continuum mechanics.

In order to compare the microscopic theory with the phenomenological
description of the MA in Sec.~\ref{macroscopic_model}, we consider
the dependence of $F_M$ on $\bm{m}$ with respect to the reference
direction $\bm{m}_\text{ref}$=[100]. Accordingly, we write the
anisotropic part $\Delta F_{M,\text{int}}$ of the intrinsic
contribution $F_{M,\text{int}}$ as
\begin{equation}
\Delta F_{M,\text{int}} =
F_{M,\text{c}}(\varepsilon_{zz},\bm{m}) -
F_{M,\text{c}}(\varepsilon_{zz},\bm{m}=[100]).
\label{F_diff_mic}
\end{equation}
In terms of the anisotropy parameters from Eq.~(\ref{F_MA_int}),
$\Delta F_{M,\text{int}}$ reads as
\begin{equation}
\Delta F_{M,\text{int}} = B_{2\perp}m_z^2 +
B_{4\parallel}(m_x^4+m_y^4-1) + B_{4\perp}m_z^4.
\label{F_diff_mac}
\end{equation}
For $\bm{m}$ rotated in the (001) and the (010) plane,
Eq.~(\ref{F_diff_mac}) can be rewritten as
\begin{equation}
\Delta F_{M,\text{int}}(\varphi)= B_{4\parallel}(\cos^4\varphi
+\sin^4\varphi-1)
\label{F_diff_phi}
\end{equation}
and
\begin{equation}
\Delta F_{M,\text{int}}(\theta)= B_{2\perp}\,\cos^2\theta
+B_{4\parallel}(\sin^4\theta -1) +B_{4\perp}\,\cos^4\theta ,
\label{F_diff_theta}
\end{equation}
respectively, where we have introduced the azimuth angle $\varphi$
and the polar angle $\theta$ with $m_x$=$\sin\theta\cos\varphi$,
$m_y$=$\sin\theta\sin\varphi$, and $m_z$=$\cos\theta$.
We proceed by calculating $\Delta F_{M,\text{int}}$ numerically
in the microscopic model as a function of $\varphi$ and $\theta$ with
$\varepsilon_{zz}$ varied in the range $-0.4\%\leq\varepsilon_{zz}\leq 0.3\%$,
using typical values for $p$, $B_{\text{G}}$, and $M$.
Equations~(\ref{F_diff_phi}) and (\ref{F_diff_theta}) are then fitted to
the resulting angular dependences using $B_{2\perp}$, $B_{4\parallel}$,
and $B_{4\perp}$ as fit parameters. For the hole density we use the
value $p$=$3.5\times10^{20}$\;cm$^{-3}$, for the exchange-splitting
parameter $B_{\text{G}}$=$-23$\;meV, and for the magnetization
$\mu_0M$=40\;mT. Inserted into Eq.~(\ref{B_G}), the latter two values
yield $A_F N_0\beta$=$-1.8$\;eV, in good agreement with the parameters
used in Ref.~\onlinecite{Die01}.

\begin{figure}[h]
\centering
\includegraphics[scale=1.3]{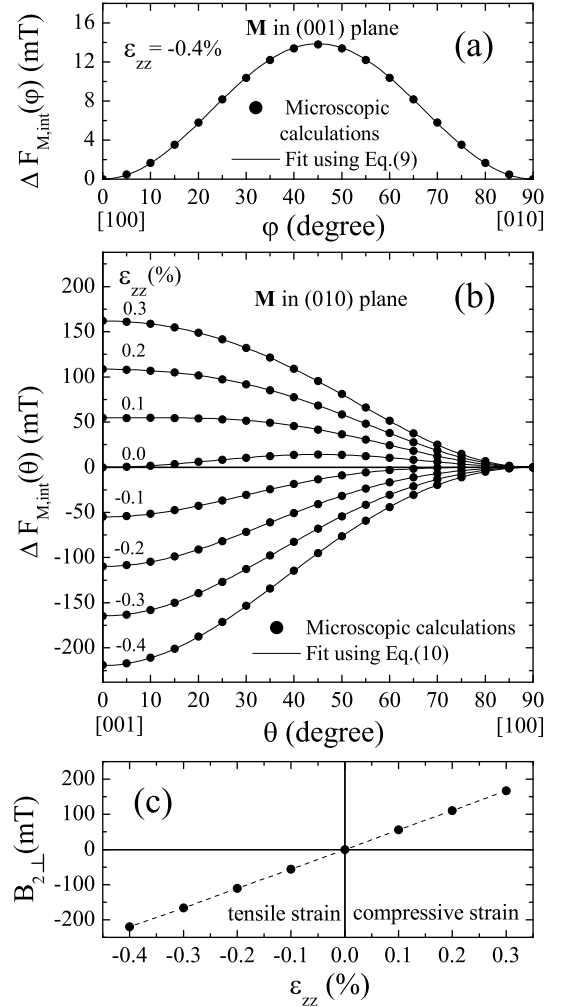}
\caption{$\Delta F_{M,\text{int}}$ calculated as a function of the
magnetization orientation and the strain within the microscopic model
(solid symbols) for $\bm{M}$ (a) in the (001) plane and (b) in the
(010) plane. $\varphi$ and $\theta$ denote the azimuth and polar
angles of $\bm{M}$, respectively. The solid lines are least-squares
fit curves using (a) Eq.~(\ref{F_diff_phi}) and (b) Eq.~(\ref{F_diff_theta})
with $B_{2\perp}$, $B_{4\parallel}$, and $B_{4\perp}$ as fit
parameters.
(c) The anisotropy parameter $B_{2\perp}$ obtained from the fit
shows a pronounced linear dependence on $\varepsilon_{zz}$.}
\label{Fit_theta}
\end{figure}
The results of the microscopic calculations are depicted by the
solid symbols in Figs.~\ref{Fit_theta}(a) and \ref{Fit_theta}(b).
Since the variation of $\Delta F_{M,\text{int}}(\varphi)$ with
$\varepsilon_{zz}$ is found to be marginal, only one
representative curve calculated with $\varepsilon_{zz}$=$-0.4\%$ is
shown in Fig.~\ref{Fit_theta}(a). It clearly reflects the fourfold
symmetry of $\Delta F_{M,\text{int}}$ within the (001) plane. By
contrast, $\Delta F_{M,\text{int}}(\theta)$ in Fig.~\ref{Fit_theta}(b)
strongly depends on $\varepsilon_{zz}$. The solid lines in
Figs.~\ref{Fit_theta}(a) and \ref{Fit_theta}(b) are least-squares
fits to the calculated data using Eqs.~(\ref{F_diff_phi}) and
(\ref{F_diff_theta}), respectively. The perfect agreement between
the microscopic results and the fit curves demonstrates that the
intrinsic part of the free energy calculated within the microscopic
theory can be well parameterized by $B_{2\perp}$, $B_{4\parallel}$,
and $B_{4\perp}$. It thus justifies the phenomenological approach
in Eq.~(\ref{F_MA_int}), taking into account only terms up to the
fourth order in $\bm{m}$. Whereas the cubic anisotropy parameters
$B_{4\parallel}$$\approx$$B_{4\perp}$$\approx$$-30$\;mT obtained from
the fit are not substantially affected by $\varepsilon_{zz}$, the
uniaxial parameter $B_{2\perp}$ exhibits a pronounced linear
dependence on $\varepsilon_{zz}$, as shown in Fig.~\ref{Fit_theta}(c).
The slope of $B_{2\perp}$($\varepsilon_{zz}$) is strongly influenced
by the exchange-splitting parameter $B_{\text{G}}$ and the hole
density $p$, as will be discussed in more detail in
Sec.~\ref{anisotropy_parameters}, Fig.~\ref{B2}(b).

The microscopic calculations show that for $\varepsilon_{zz}<0$
(tensile strain) $\Delta F_{M,\text{int}}$ exhibits two equivalent
minima for $\bm{m}$ oriented along [001] and [00$\bar{1}$], which
become more pronounced with increasing tensile strain. In contrast,
in the regime of compressive strain ($\varepsilon_{zz}>0$), the minima
occur for $\bm{m}$ along [100], [$\bar{1}$00], [010], and [0$\bar{1}$0].
Thus, the theoretical model correctly describes the well known
experimental fact that for sufficiently high hole densities and low
temperatures the magnetically hard axis along [001] in compressively
strained layers turns into an easy axis in tensily strained layers.
Note, however, that for a quantitative comparison between experiment
and theory the extrinsic contributions to $\Delta F_{M}$ have also
to be taken into account.

\section{\label{results}Results and discussion}

In the following, the experimental data obtained for the (Ga,Mn)As
samples under study are discussed.

\subsection{\label{structural}Lattice parameters and strain}

\begin{figure}[h]
\includegraphics[scale=0.9]{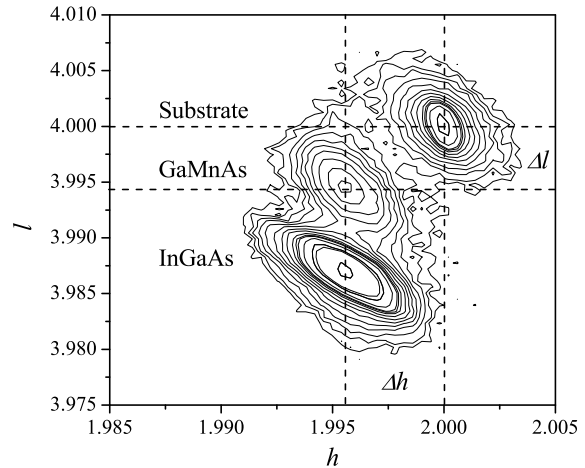}
\caption{Reciprocal space map around the asymmetric (224) reflections
of a nearly unstrained (Ga,Mn)As layer ($\varepsilon_{zz}$=$-0.04$\%)
grown on relaxed (In,Ga)As/GaAs template.
$h$ and $l$ are the coordinates in $k$-space in units of the reciprocal
lattice vectors along [100] and [001] in the GaAs substrate,
respectively. $\Delta h$ and $\Delta l$ denote the shift of the peak
position relative to that of the substrate. In the plot, $\Delta h$ and
$\Delta l$ are only depicted for the (Ga,Mn)As layer.}
\label{RSM}
\end{figure}

\begin{figure}[h]
\begin{center}
\includegraphics[scale=0.8]{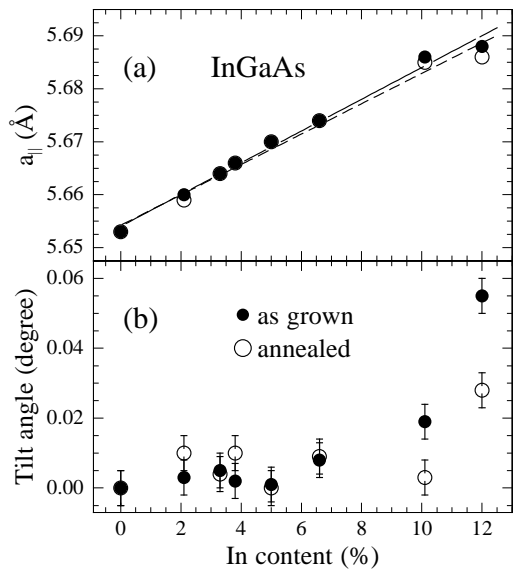}
\caption{\label{tilt}
(a) Lateral lattice parameter $a_\parallel$ and (b) tilt angle towards
the [110] direction of the (In,Ga)As buffer layer, plotted against the
In content. The solid line (as grown) and the dashed line (annealed) are
regression lines.}
\end{center}
\end{figure}

In order to study the structural properties of the
(Ga,Mn)As/(In,Ga)As/GaAs samples, reciprocal space maps (RSM)
of the asymmetric (224), ($\bar{2}$$\bar{2}$4), ($\bar{2}$24), and
(2$\bar{2}$4) reflections were recorded using HRXRD. Figure~\ref{RSM}
exemplarily shows an RSM contour plot of the (224) reflection for a
nearly unstrained (Ga,Mn)As layer with $\varepsilon_{zz}$=$-0.04$\%,
depicting separate peaks for the GaAs substrate, the (In,Ga)As buffer,
and the (Ga,Mn)As layer.
From the peak positions ($h$,$l$) and their shifts ($\Delta h$,$\Delta l$)
relative to that of the substrate the lateral and vertical lattice
parameters of (Ga,Mn)As and (In,Ga)As can be determined using the
relations
\begin{equation}
a_\parallel=a_{\rm s}(1-\Delta h/h),\quad  a_\perp=a_{\rm s}(1-\Delta l/l),
\label{aparaperp}
\end{equation}
where $a_{\text{s}}$ denotes the lattice constant of the GaAs
substrate. $h$ and $l$ are the coordinates in $k$-space referring to
the reciprocal lattice vectors of GaAs along the [100] and [001]
directions, respectively. In Fig.~\ref{RSM}, the peaks of (Ga,Mn)As
and (In,Ga)As are centered at the same value of $h$, confirming that
the (Ga,Mn)As layer has been grown lattice matched to the (In,Ga)As
buffer. The lateral lattice parameters $a_\parallel$ are therefore
the same in the (In,Ga)As and (Ga,Mn)As layers. The shift $\Delta h$
from the substrate peak at $h$=2 to lower values is due to the strain
relaxation in the buffer layer ($a_\parallel$$>$$a_{\text{s}}$). The
lattice parameters $a_\parallel$ of the (In,Ga)As templates in the
as-grown and annealed samples are plotted against the In content
in Fig.~\ref{tilt}(a). Apparently, post-growth annealing had no
significant influence on $a_\parallel$, which linearly increases
with the In content.

For both (Ga,Mn)As and (In,Ga)As, the HRXRD measurements yielded
different values of ($h$,$l$) for the (224) and ($\bar{2}$$\bar{2}$4)
reflections, revealing a tilt of the lattice towards the [110]
direction.\cite{Kro96} The tilt of the (Ga,Mn)As layer originates
from an equal tilt in the (In,Ga)As buffer pointing to an anisotropic
relaxation of the (In,Ga)As templates, typically found in layers
grown on vicinal substrates.\cite{Aye91} The samples under
investigation, however, were grown on non-miscut (001) wafers. As can
be seen in Fig.~\ref{tilt}(b), the measured tilt angles tend to higher
values with increasing In fraction.
It should be emphasized that it is imperative to take the tilt
into account when determining the lattice parameters in order to avoid
erroneous results. This can be done by inserting into
Eq.~(\ref{aparaperp}) the averaged values of the peak positions and
shifts obtained from the (224) and ($\bar{2}$$\bar{2}$4) reflections.
For the angle-dependent magnetotransport measurements, the influence
of the tilt is negligible since the tilt angles observed are smaller
than 0.06 degree.
\begin{figure}[h]
\begin{center}
\includegraphics[scale=0.8]{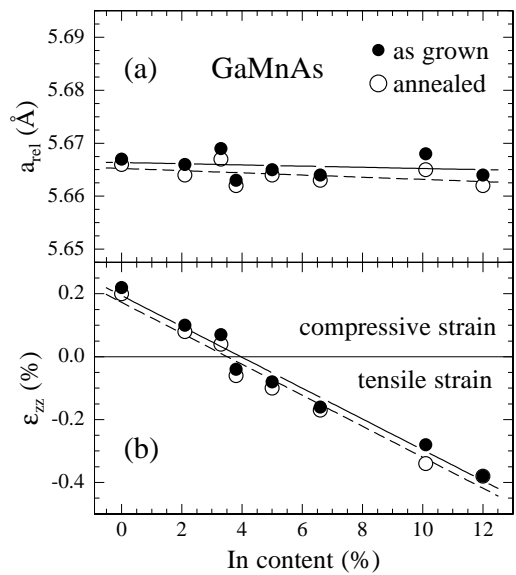}
\caption{\label{ezz} (a) Relaxed lattice parameter $a_{\text{rel}}$
and (b) vertical strain $\varepsilon_{zz}$ of the (Ga,Mn)As layer
plotted against the In content in the (In,Ga)As buffer. Post-growth
annealing only leads to a slight decrease of the values.}
\end{center}
\end{figure}

The relaxed lattice parameter $a_{\text{rel}}$ of a biaxially strained
layer on (001)-oriented substrate is obtained from the relation
\begin{equation}
a_{\rm{rel}} = \frac{2C_{12}}{C_{11}+2C_{12}}a_\parallel +
\frac{C_{11}}{C_{11}+2C_{12}}a_\perp\;,
\end{equation}
where $C_{11}$ and $C_{12}$ are elastic stiffness constants. As an
approximation, we use the values $C_{11}$=11.90$\times10^{10}$\;Pa
and $C_{12}$=5.34$\times10^{10}$\;Pa of GaAs for both the (Ga,Mn)As
and the (In,Ga)As layers.\cite{Lev96} In Fig.~\ref{ezz}(a),
$a_{\text{rel}}$ is shown for the (Ga,Mn)As layers as a function of
the In content in the (In,Ga)As buffer. It is found to be nearly
unaffected by the (In,Ga)As template underneath. The fluctuations
in the values of $a_{\text{rel}}$ are mainly attributed to slight
variations of the growth temperature. For all (In,Ga)As templates
under study, the degree of relaxation defined by
$R$=$(a_{\parallel}-a_{\text{s}})/(a_{\text{rel}}-a_{\text{s}})$
was above 80\%. Once the vertical and relaxed lattice parameters of
the (Ga,Mn)As layers are known, the vertical strain $\varepsilon_{zz}$
can be calculated from the relation
\begin{equation}
\varepsilon_{zz}=(a_\perp-a_{\rm{rel}})/a_{\rm{rel}}.
\label{varezz}
\end{equation}
In Fig.~\ref{ezz}(b), $\varepsilon_{zz}$ is plotted against the In
content. The slight decrease of $a_{\text{rel}}$ and $\varepsilon_{zz}$
upon annealing is supposed to arise from the outdiffusion and/or
rearrangement of the highly mobile
Mn$_{\text{I}}$.\cite{Pot01,Yu02,Edm04,Zha05,Lim05}

\subsection{\label{Sec_p_TC} Hole density and Curie temperature}

\begin{figure}[h]
\includegraphics[scale=0.8]{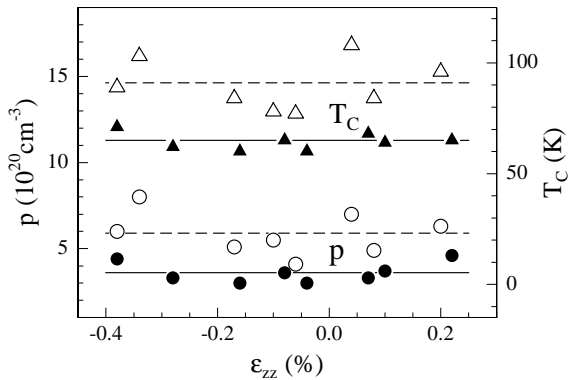}
\caption{\label{a_p_tc} Hole density $p$ and Curie temperature
$T_{\text{C}}$ of (Ga,Mn)As plotted against the strain
$\varepsilon_{zz}$ for the as-grown (solid symbols) and the annealed
samples (open symbols). The fluctuations in the values are mainly
attributed to slight variations of the growth temperature.
For $p$ we estimate an error margin of about $\pm$10\% and
for $T_{\text{C}}$ an error margin of up to $\pm$20\%.}
\end{figure}

Determination of the hole concentrations $p$ in (Ga,Mn)As is
complicated by a dominant anomalous contribution to the Hall
effect proportional to the normal component of the magnetization
$\bm{M}$ (anomalous Hall effect). To overcome this problem,
magnetotransport measurements were performed at high magnetic
fields up to 14.5\;T.
Assuming the magnetization to be saturated perpendicular
to the layer plane at magnetic fields $\mu_0H$$\gtrsim$4\;T,
the measured transverse resistivity was fitted using the equation
\begin{equation}
\rho_{\text{trans}}(H) = R_0 \mu_0H + c_1\rho_{\text{long}}(H)
+ c_2 \rho_{\text{long}}^2(H)
\end{equation}
for the ordinary and anomalous Hall effect with $R_0$, $c_1$, and
$c_2$ as fit parameters. Here $R_0$=$1/ep$ is the ordinary Hall
coefficient and $\rho_{\text{long}}$ the measured field-dependent
longitudinal resistivity.
The second term on the right hand side arises from skew
scattering\cite{Smi55,Smi58} and the third term from
side jump scattering\cite{Ber70} and/or Berry phase
effects.\cite{Jun02} As mentioned in Sec.~\ref{experimental}, the
Curie temperatures $T_{\text{C}}$
were inferred from the peak positions of the temperature-dependent
resistivities $\rho_{\text{long}}$.\cite{Esc97,Mat98} Considering that
the $T_{\text{C}}$ values thus obtained generally differ from those
determined by temperature-dependent magnetization
measurements,\cite{Nov08} we estimate an error margin of up to 20\%.
Similar to $a_{\text{rel}}$, neither $p$ nor $T_{\text{C}}$ are
significantly influenced by the strain as shown in Figure~\ref{a_p_tc}.
The values scatter around
$p_{\text{ag}}$=3.5$\times10^{20}$\;cm$^{-3}$ and
$T_{\text{C}}$=65\;K for the as-grown samples and
$p_{\text{ann}}$=5.8$\times 10^{20}$\;cm$^{-3}$ and
$T_{\text{C}}$=91\;K for the annealed samples.
It is well known that the lattice constant, the hole density,
and the Curie temperature strongly depend on the concentration of
Mn$_{\text{Ga}}$ acceptors, Mn$_{\text{I}}$ double donors, and
other compensating defects such as As$_{\text{Ga}}$ antisites. The
insensitivity of $a_{\text{rel}}$, $p$, and $T_{\text{C}}$ to strain
in the as-grown and annealed samples under study suggests the assumption
that strain has no significant influence on the incorporation of
Mn$_{\text{Ga}}$, Mn$_{\text{I}}$, and As$_{\text{Ga}}$. At least the
sum of the changes caused by the different constituents seems to be
unaltered. Moreover, the insensitivity of $T_{\text{C}}$ with respect
to $\varepsilon_{zz}$ supports
theoretical predictions that the magnetic coupling should be
unaffected by strain, since the corresponding deformation energies
are expected to be too small to significantly enhance or reduce
the p-d kinetic exchange interaction.\cite{Die01,Jun06}

\subsection{Anisotropy parameters \label{anisotropy_parameters}}

\begin{figure}[h]
\includegraphics[scale=0.40]{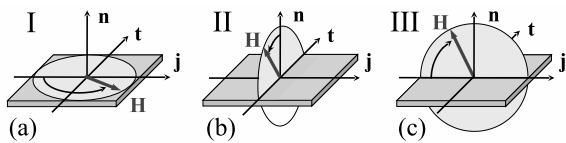}
\caption{\label{planes} The angular dependence of the resistivities
was probed by rotating an external magnetic field $\bm{H}$ within
the three different planes (a) perpendicular to $\bm{n}$, (b)
perpendicular to $\bm{j}$, and (c) perpendicular to $\bm{t}$.
The corresponding configurations are referred to as I, II, and
III.}
\end{figure}

\begin{figure}[htb]
\centering
\includegraphics[scale=1.0]{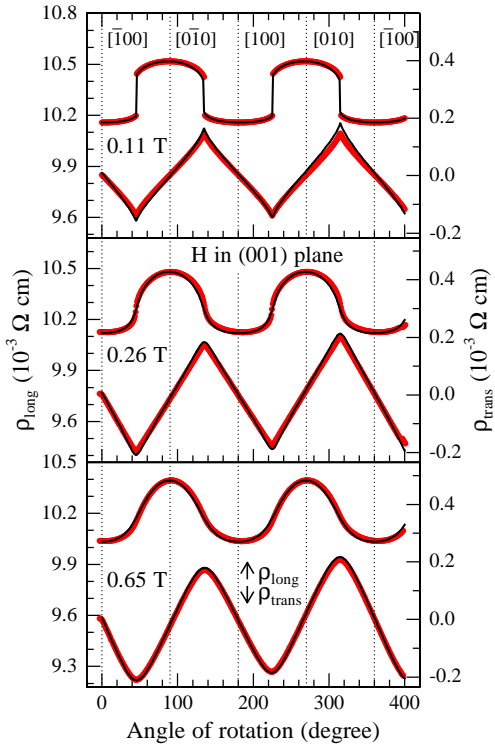}
\caption{(Color online) Resistivities $\rho_{\text{long}}$ and
$\rho_{\text{trans}}$ recorded from a nearly unstrained (Ga,Mn)As layer
with $\varepsilon_{zz}$=$-0.04$\% at 4.2\;K and $\bm{j}\parallel [100]$
(red solid circles). The measurements were performed at fixed field
strengths of $\mu_0H$=0.11, 0.26 and 0.65\;T with $\bm{H}$ rotated in
the (001) plane corresponding to configuration I. The black solid lines
are fits to the experimental data using Eqs.~(\ref{rho_long}) and
(\ref{rho_trans}), and one single set of resistivity and anisotropy
parameters.} \label{magnetotransport}
\end{figure}

Experimental values for the anisotropy parameters
$B_{001}$=$B_{2\perp}$+$B_{\text{d}}$,
$B_{4\parallel}$, $B_{4\perp}$, and $B_{\bar110}$ were determined
by means of angle-dependent magnetotransport measurements. A detailed
description of the corresponding procedure is given in the
Refs.~\onlinecite{Lim06a} and \onlinecite{Lim08}. It can be briefly
summarized as follows. The longitudinal and transverse resistivities
$\rho_{\text{long}}$ and $\rho_{\text{trans}}$, respectively, are
measured as a function of the magnetic field orientation at fixed
field strengths of $\mu_0H$=0.11, 0.26, and 0.65\;T. At each field
strength, $\bm{H}$ is rotated within three different crystallographic
planes perpendicular to the directions $\bm{n}$, $\bm{j}$, and
$\bm{t}$, respectively. The corresponding configurations, labeled I,
II, and III, are shown in Fig.~\ref{planes}.
The vectors form a right-handed coordinate system, where $\bm{j}$
defines the current direction, $\bm{n}$ the surface normal, and
$\bm{t}$ the transverse direction. For current directions
$\bm{j} \parallel [100]$ and $\bm{j} \parallel [110]$, the
resistivities can be written as\cite{Lim08}
\begin{equation}
 \rho_{\text{long}} = \rho_0 + \rho_1 m_j^2 + \rho_2 m_n^2 +
 \rho_3 m_j^4 + \rho_4 m_n^4 + \;\rho_5 m_j^2m_n^2 ,
 \label{rho_long}
\end{equation}
\begin{equation}
\rho_{\text{trans}} = \rho_6 m_n + \rho_7 m_jm_t + \rho_8 m_n^3 +
 \rho_9 m_jm_tm_n^2,
\label{rho_trans}
\end{equation}
where $m_j$, $m_t$, and $m_n$ denote the components of $\bm{m}$ along
$\bm{j}$, $\bm{t}$, and $\bm{n}$, respectively. At sufficiently high
magnetic fields, the Zeeman energy in $G_M(\bm{m})$ dominates and the
magnetization direction $\bm{m}$ follows the orientation $\bm{h}$ of
the external field. The resistivity parameters $\rho_i$ ($i$=1,...,9)
are then obtained from a fit of the Eqs.~(\ref{rho_long}) and
(\ref{rho_trans}) to the experimental data recorded at 0.65\;T. With
decreasing field strength, the influence of the MA increases and
$\bm{m}$ more and more deviates from $\bm{h}$. Controlled by the
magnetic anisotropy parameters, the shape of the measured resistivity
curves changes and $B_{001}$, $B_{4\parallel}$, $B_{4\perp}$, and
$B_{\bar110}$ are obtained from a fit to the data recorded at
$\mu_0H$=0.26 and 0.11\;T. In the fit procedure, $\bm{m}$ is calculated
for every given magnetic field $\bm{H}$ by numerically minimizing
$G_{M}$ with respect to $\bm{m}$. Figure~\ref{magnetotransport}
exemplarily shows the angular dependence of $\rho_{\text{long}}$ and
$\rho_{\text{trans}}$ for a nearly unstrained (Ga,Mn)As layer
($\varepsilon_{zz}$=$-0.04$\%) with $\bm{H}$ rotated in the (001)
plane (configuration I) and $\bm{j}\parallel [100]$. The experimental
data are depicted by red solid circles and the fits by black solid
lines.

\begin{figure}[h]
\begin{center}
\includegraphics[scale=1.0]{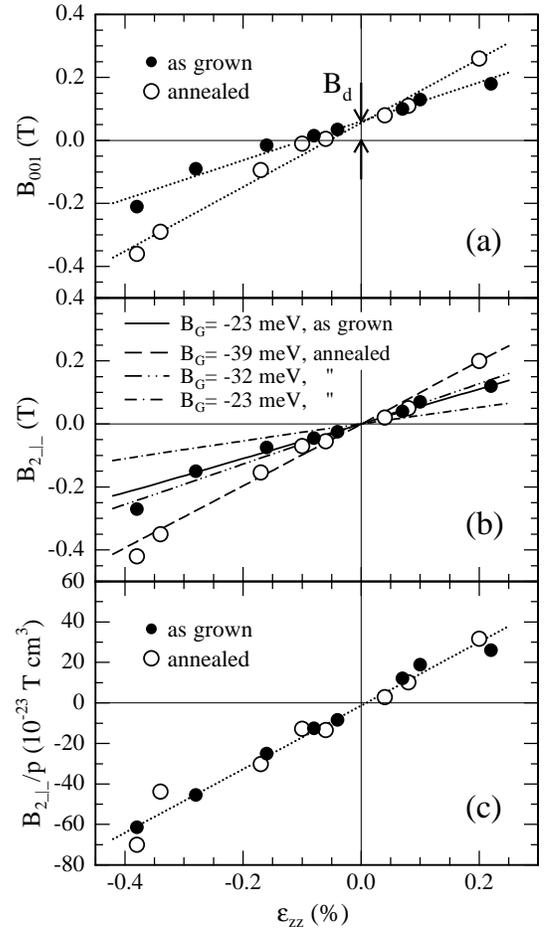}
\caption{\label{B001} (a) Dependence of the uniaxial out-of-plane
anisotropy parameter $B_{001}$=$B_{2\perp}$+$B_{\text{d}}$ on
$\varepsilon_{zz}$ for the as-grown (solid circles) and annealed samples
(open circles). $B_{\text{d}}$=60\;mT is inferred from the intersections
between the corresponding regression lines (dotted lines) and the vertical
axis. (b) Anisotropy parameter $B_{2\perp}$ obtained by subtracting 60\;mT
from the measured $B_{001}$ data. The lines are model calculations for
$B_{2\perp}$ performed within the microscopic theory described in
Sec.~\ref{comparison} using values for the parameter $B_{\text{G}}$ as
shown and the averaged $p$ values from Fig.~\ref{a_p_tc}.
(c) Anisotropy parameter $B_{2\perp}$=$B_{001}-B_{\text{d}}$ normalized
to the hole density $p$. Experimentally, the same linear dependence of
$B_{2\perp}/p$ on $\varepsilon_{zz}$ is obtained for both the as-grown
and the annealed samples. The dotted line represents a linear regression.}
\label{B2}
\end{center}
\end{figure}

\begin{figure}[h]
\begin{center}
\includegraphics[scale=1.0]{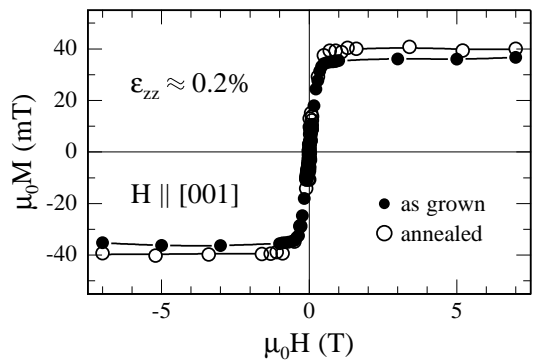}
\caption{\label{squid} SQUID curves of the as-grown and annealed samples
with $\varepsilon_{zz}$$\approx$0.2\%, measured at 5\;K for $\bm{H}$
oriented along [001]. The saturation magnetization of $\sim$40\;mT is
representative for the whole set of (Ga,Mn)As samples.
The values of the coercive fields, derived from the hysteresis
loops (not shown), were found to be below 10 mT.}
\end{center}
\end{figure}

\begin{figure}[h]
\begin{center}
\includegraphics[scale=1.0]{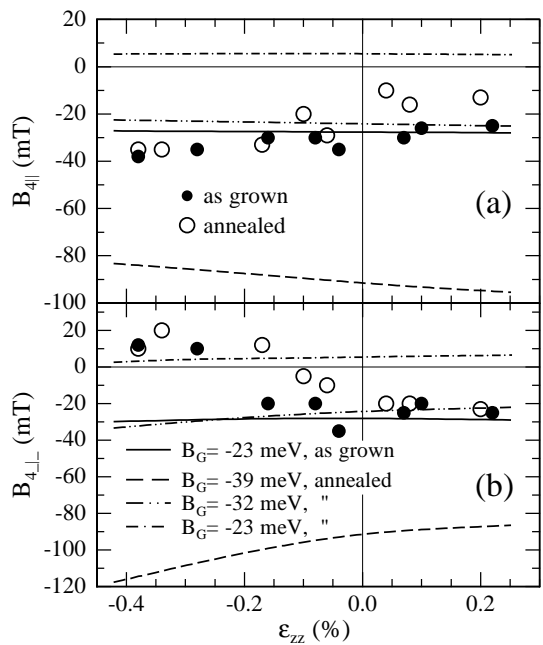}
\caption{\label{B4} Dependence of the cubic anisotropy parameters
(a) $B_{4\parallel}$ and (b) $B_{4\perp}$ on $\varepsilon_{zz}$
for the as-grown (solid circles) and annealed samples (open circles).
The lines are model calculations within the microscopic theory using the
same values for $p$ and $B_{\text{G}}$ as in Fig.~\ref{B001}(b).}
\end{center}
\end{figure}

\begin{figure}[h]
\begin{center}
\includegraphics[scale=1.0]{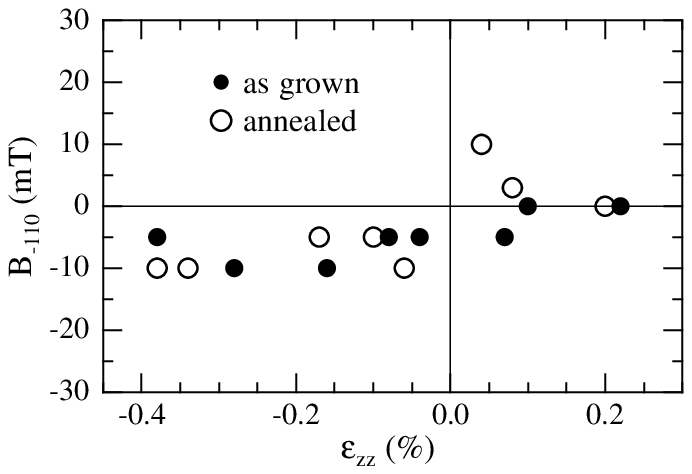}
\caption{\label{Bm110} Dependence of the uniaxial in-plane parameter
$B_{\bar110}$ for the as-grown (solid circles) and annealed samples
(open circles).}
\end{center}
\end{figure}

Applying the procedure described above to the whole set of (Ga,Mn)As
layers under study, the resistivity parameters $\rho_i$ ($i$=1,...,9)
and the anisotropy parameters $B_{001}$=$B_{2\perp}$+$B_{\text{d}}$,
$B_{4\parallel}$, $B_{4\perp}$, and $B_{\bar110}$ were determined as
functions of the vertical strain $\varepsilon_{zz}$. The results for
the resistivity parameters were extensively discussed in
Ref.~\onlinecite{Lim08}. In the present work, we exclusively focus on
the anisotropy parameters.

Figure~\ref{B001}(a) shows the values of the parameter
$B_{001}$=$B_{2\perp}$+$B_{\text{d}}$ describing the uniaxial
out-of-plane anisotropy. For both the as-grown and the annealed
samples, a pronounced linear dependence on $\varepsilon_{zz}$ is found
in qualitative agreement with the microscopic model calculations
presented in Fig.~\ref{Fit_theta}(c). For zero strain, the cubic
symmetry requires $B_{2\perp}$=0 and the extrinsic parameter
$B_{\text{d}}$$\approx$60\;mT is inferred from the intersections
between the regression lines (dotted lines) and the vertical axis.
If shape anisotropy was the only extrinsic contribution to the $m_z^2$
term of $F_M$, as assumed in Sec.~\ref{macroscopic_model}, the value
$B_{\text{d}}$$\approx$60\;mT would correspond to a sample magnetization
of $\mu_0 M$=$2B_{\text{d}}$$\approx$120\;mT. This value, however, exceeds
the saturation magnetization determined by SQUID measurements by a
factor of $\sim$3. Figure~\ref{squid} shows as an example the SQUID
curves obtained for the as-grown and annealed samples with
$\varepsilon_{zz}$$\approx$0.2\%. At the moment, the reason for
this discrepancy is not yet understood. We suspect, however, that it
might be related to the vertical gradient in the MA mentioned at the
end of Sec.~\ref{experimental}. Assuming the value of $B_{\text{d}}$
to be nearly the same for all samples under consideration, the
strain-dependent intrinsic parameter $B_{2\perp}$ is obtained by
subtracting $B_{\text{d}}$ from the measured $B_{001}$ data. In
Fig.~\ref{B001}(b), the values of $B_{2\perp}$ derived in this way
are shown together with model calculations performed within the
microscopic theory (see Sec.~\ref{comparison}). Using for $p$ the
mean values $p_{\text{ag}}$=3.5$\times 10^{20}$\;cm$^{-3}$ and
$p_{\text{ann}}$=5.8$\times 10^{20}$\;cm$^{-3}$ (see Sec.~\ref{Sec_p_TC}),
the calculated values of $B_{2\perp}$ are found to be in quantitative
agreement with the experimental data if $B_{\text{G}}$=$-23$\;meV is
chosen for the as-grown samples and $B_{\text{G}}$=$-39$\;meV for the
annealed samples. The corresponding curves are depicted by the solid and
dashed lines in Fig.~\ref{B001}(b). Comparing the exchange-splitting
parameters with the respective hole densities, we find
\begin{equation}
\frac{B_{\text{G,ann}}}{B_{\text{G,ag}}} \approx
\frac{p_{\text{ann}}}{p_{\text{ag}}} \approx 1.7\,.
\label{BGp}
\end{equation}
For the annealed samples, theoretical
results are also shown for $-32$\;meV  and $-23$\;meV, demonstrating
that the slope of $B_{2\perp}$($\varepsilon_{zz}$) is drastically
reduced with decreasing $B_{\text{G}}$. Remarkably, normalization of
the experimentally derived values of $B_{2\perp}$ to the corresponding
hole concentration $p$ yields the same linear dependence of
$B_{2\perp}/p$ on $\varepsilon_{zz}$ for both the as-grown and the
annealed samples as shown in Fig.~\ref{B001}(c). We thus find, at
least for the range of hole densities and strain under consideration,
the experimental relationship
\begin{equation}
B_{2\perp} = K p\varepsilon_{zz}\,,
\label{B2pe}
\end{equation}
with $K$=1.57$\times$10$^{-19}$\;T\,cm$^3$. If we assume a linear relation
$B_{\text{G}}$=$A_\text{F}\beta M/6g\mu_\text{B} \propto p$, in accordance
with Eq.~(\ref{BGp}), Eq.~(\ref{B2pe}) can be reproduced by the microscopic
theory. Note however, that this is not trivial, since
$\Delta F_{M,\text{int}}$ explicitly depends on both $B_{\text{G}}$
and $p$. Anyhow, the relation $B_{\text{G}} \propto p$ demands a future
detailed investigation.

The experimental values of the fourth-order parameters $B_{4\parallel}$
and $B_{4\perp}$ are presented in Figs.~\ref{B4}(a) and \ref{B4}(b),
respectively. Whereas $B_{4\parallel}$ only slightly varies between $-40$
and $-10$\;mT, $B_{4\perp}$ exhibits positive values close to 10\;mT for
$\varepsilon_{zz}$$\lesssim$$-0.15$\% and negative values of about
$-20$\;mT for $\varepsilon_{zz}$$\gtrsim$$-0.15$\%. The lines depicted
in Fig.~\ref{B4} were calculated using the same values for $p$ and
$B_{\text{G}}$ as in Fig.~\ref{B001}(b).
Obviously, the change of sign of $B_{4\perp}$ does not appear in the
theoretical curves. Apart from this disagreement,
the experimental data of the as-grown samples are again well reproduced
for $B_{\text{G}}$=$-23$\;meV. In the case of the annealed samples, now
the splitting parameter $B_{\text{G}}$=$-32$\;meV yields a much better
description of the measured data than $B_{\text{G}}$=$-39$\;meV, found to
fit the strain dependence of $B_{2\perp}$.

In view of the perfect quantitative interpretation of $B_{2\perp}$
by the mean-field Zener model, this small discrepancy and the fact that
the change of sign of $B_{4\perp}$ is not reproduced by the calculations,
should not be overestimated. The experimentally observed change of sign
of $B_{4\perp}$ may be caused by extrinsic influences, not accessible by
the model, such as the increasing density of threading dislocations in the
(Ga,Mn)As/(In,Ga)As layers with increasing In concentration. Moreover, we
cannot rule out that the change of sign is an artifact of the experimental
method for determining the relatively small anisotropy parameter $B_{4\perp}$.
The same problem has also been reported in Ref.~\onlinecite{Liu03}, where
the authors did not extract information on the fourth-order out-of-plane
anisotropy parameters from their ferromagnetic resonance measurements,
since the corresponding contributions to the MA were masked by the much
larger contributions of the second-order out-of-plane term. As to the
differing values of $-39$\;meV and $-32$\;meV for $B_{\text{G}}$ in the
case of the annealed samples,
it should be pointed out that for high hole densities the Fermi energy
shifts deep into the valence band.
Therefore, the values of the free energy obtained by using a
6$\times$6 $\bm{k}\cdot\bm{p}$ effective Hamiltonian become increasingly
unreliable when analyzing higher order contributions to the free energy.
One should not jump to the conclusion that this has to be interpreted
as a deficiency of the Zener model itself.

In agreement with the data presented in Refs.~\onlinecite{Liu03},
\onlinecite{Wan05}, and \onlinecite{Yam06}, the experimental values of
the extrinsic uniaxial in-plane parameter $B_{\bar110}$ are much smaller
than those of the cubic in-plane parameter $B_{4\parallel}$ at 4.2\;K.
As shown in Fig.~\ref{Bm110}, they scatter around zero with
$-10$\;mT$\leq$$B_{\bar110}$$\leq$10\;mT.

Due to the strong dependence of $B_{001}$ on $\varepsilon_{zz}$,
the out-of-plane axis [001] becomes magnetically harder with
$\varepsilon_{zz}$ increasing from $-0.4$\% to 0.2\%. Neglecting the
small influence of $B_{\bar110}$ in our samples at 4.2\;K, the critical
strain $\varepsilon_{zz}^{\text{crit}}$, where a reorientation of the
easy axis from out-of-plane to in-plane occurs, can be estimated from
the condition $B_{001}$+$B_{4\perp}$=$B_{4\parallel}$. We obtain for
the as-grown (Ga,Mn)As layers under study
$\varepsilon_{zz}^{\text{crit}}$=$-0.13$\% and for the annealed layers
$\varepsilon_{zz}^{\text{crit}}$=$-0.07$\%. Remarkably, these values
are very close to the $\varepsilon_{zz}$ values in Fig.~\ref{B4} where
$B_{4\perp}$ changes sign.

Since the MA sensitively depends on the individual growth conditions,
care has to be taken when comparing the values of anisotropy parameters
published by different groups. Keeping this restriction in mind, the
data presented in this work are in reasonable agreement, e.g., with
the results obtained by Liu et al.\cite{Liu03} for a representative
pair of compressively and tensily strained (Ga,Mn)As samples with
3\% Mn.

\subsection{Magnetostriction constant \label{magnetostriction_constant}}

Changing the magnetization of a ferromagnet, e.g. by an external
magnetic field, leads to a variation of its geometrical shape. For
crystals with cubic symmetry, the relative elongation $\lambda$ in a given
direction $\bm{\beta}$ can be expressed in terms of the magnetization
orientation $\bm{m}$ and the magnetostriction constants $\lambda_{100}$
and $\lambda_{111}$ along [100] and [111], respectively, according to
\cite{Von74}
\begin{eqnarray}
\lambda &=& \frac{3}{2}\lambda_{100}
\left(m_x^2\beta_x^2+m_y^2\beta_y^2+m_z^2\beta_z^2-\frac{1}{3}\right)
\nonumber \\
&& +3\lambda_{111}\left(m_x m_y \beta_x \beta_y+m_y m_z \beta_y
\beta_z+m_x m_z \beta_x \beta_z\right)\;.
\end{eqnarray}
Starting from Eq.~(\ref{B2pe}), we are able to determine $\lambda_{100}$
defined by \cite{Von74}
\begin{equation}
\lambda_{100}=\frac{2}{9}\frac{a_1}{C_{12}-C_{11}}\,,
\label{magnetostriction}
\end{equation}
where $a_1$ denotes the magnetoelastic coupling constant.
First-order expansion of the free-energy density
$F(\varepsilon_{ij},\bm{m})$ with respect to $\varepsilon_{ij}$
and continuum mechanics yield the relation
$B_{2\perp}M$=$\varepsilon_{zz}a_1(1+C_{11}/2C_{12})$. Thus,
Eq.~(\ref{magnetostriction}) can be rewritten as
\begin{equation}
\lambda_{100}=\frac{2KMp}{9(C_{12}-C_{11})(1+C_{11}/2C_{12})}\,.
\end{equation}
Inserting the experimental values for $p$, $\mu_0M$, and $K$, we obtain
$\lambda_{100}$$\approx$$-3$\;ppm for the as-grown samples and
$\lambda_{100}$$\approx$$-5$\;ppm for the annealed samples. In
Ref.~\onlinecite{Bih08a}, we already deduced an approximately constant
value of $\lambda_{111}$$\approx$5\;ppm below 40\;K decreasing to zero
at higher temperatures (40\;K$<$T$<$$T_{\text{C}}$$\approx$85\;K) via
applying piezo stress along the [110] direction of a piezoelectric
actuator/(Ga,Mn)As hybrid structure. Our results for both
magnetostriction constants are close to the values
$\lambda_{100}$=$-11.3$\;ppm and $\lambda_{111}$=8.1\;ppm reported
by Masmanidis et al.\cite{Mas05}

\section{Summary}

A series of (Ga,Mn)As layers with 5\% Mn was grown on relaxed
graded (In,Ga)As/GaAs templates with In contents up to 12\%. In this
way, the vertical strain $\varepsilon_{zz}$ in the (Ga,Mn)As layers
could be gradually varied over a wide range from $-0.38$\% (tensile
strain) to 0.22\% (compressive strain). The strain was found to have
no significant influence on the hole concentration, the Curie
temperature, and the relaxed lattice parameter. Angle-dependent
magnetotransport measurements were performed to determine the uniaxial
and cubic anisotropy parameters. $B_{2\perp}$ turned out to be
proportional to both the strain and the hole concentration. From this
linear dependence, the magnetostriction constant $\lambda_{100}$ was
determined. While $B_{2\parallel}$ and $B_{4\parallel}$ are nearly
strain independent, $B_{4\perp}$ changes sign when the magnetic
easy axis flips from in-plane to out-of-plane.

Microscopic calculations of the free-energy density were
performed based on the mean-field Zener model of Dietl et al.\cite{Die01}
They justify the approximations made in the parameterization of the free
energy by considering only terms up to the fourth order. The
strain-dependent anisotropy parameters derived from the calculations
were found to be in good quantitative agreement with the experimental
results. In the case of the as-grown samples, the values of $B_{2\perp}$
and $B_{4\parallel}$ are even perfectly reproduced. The quantitative
comparison between the comprehensive set of experimental data and the
microscopic calculations may be considered a valuable contribution to
the ongoing controversy on impurity band versus valence band in
(Ga,Mn)As.\cite{Jun07,Bur08}

Using the orientation of the magnetization as the basic
information bit of a non-volatile memory, the tayloring and manipulation
of the MA is of special importance. As shown in this work, the choice
of an appropriate (In,Ga)As template allows for an adjustment of
$\varepsilon_{zz}$ close to the critical value $\varepsilon_{zz}^{\text{crit}}$.
Then, the magnetization direction can be switched from in-plane to
out-of-plane or vice versa, e.g., by minute variation of the piezo stress
in a piezoelectric actuator/(Ga,Mn)As hybrid structure\cite{Bih08a}.

\begin{acknowledgments}
This work was supported by the Deutsche Forschungsgemeinschaft
under Contract No. Li 988/4.
\end{acknowledgments}


\begin{thebibliography}{99}
\bibitem{rem}
Throughout this work, unit vectors are denoted by bold lower-case
letters.
\end{thebibliography}
\end{document}